\ifx\pdfoutput\undefined        
  \documentclass{an}
  \usepackage{url}
\else                           
  \documentclass[pdftex]{an}
  \usepackage[bookmarks=false]{hyperref}
\fi

\usepackage{times}
\usepackage{graphicx}
\usepackage{fancyhdr}
\usepackage{url}
\usepackage{bm}
\newcommand{\EQ}{\begin{equation}}
\newcommand{\EN}{\end{equation}}
\newcommand{\EQA}{\begin{eqnarray}}
\newcommand{\ENA}{\end{eqnarray}}

\newcommand{\Eq}[1]{Eq.~(\ref{#1})}

\newcommand{\Sec}[1]{Sect.~\ref{#1}}

\newcommand{\Fig}[1]{Fig.~\ref{#1}}

\newcommand{\FFigs}[2]{Figs~\ref{#1}--\ref{#2}}

\newcommand{\bra}[1]{\langle #1\rangle}

\newcommand{\meanEMF}{\overline{\vec{\cal E}}}
\newcommand{\meanemf}{\overline{\cal E}}

\newcommand{\meanB}{\overline{B}}
\newcommand{\meanJ}{\overline{J}}
\newcommand{\meanU}{\overline{U}}

\newcommand{\meanO}{\overline{\Omega}}

\newcommand{\meanBB}{\overline{\vec{B}}}
\newcommand{\meanJJ}{\overline{\vec{J}}}
\newcommand{\meanUU}{\overline{\vec{U}}}
\newcommand{\meanWW}{\overline{\vec{W}}}

\newcommand{\eee}{\hat{\mbox{\boldmath $e$}} {}}

\newcommand{\zz}{\hat{\mbox{\boldmath $z$}} {}}

\newcommand{\xx}{\mbox{\boldmath $x$} {}}

\newcommand{\ddelta}{{\vec{\delta}}}
\newcommand{\ggamma}{{\vec{\gamma}}}
\newcommand{\UU}{{\vec{U}}}
\newcommand{\uu}{{\vec{u}}}
\newcommand{\BB}{{\vec{B}}}
\newcommand{\JJ}{{\vec{J}}}

\newcommand{\AAA}{{\vec{A}}}

\newcommand{\bb}{{\vec{b}}}

\newcommand{\ff}{\mbox{\boldmath $f$} {}}

\newcommand{\GG}{\mbox{\boldmath $G$} {}}
\newcommand{\kk}{\mbox{\boldmath $k$} {}}

\newcommand{\nab}{\mbox{\boldmath $\nabla$} {}}
\newcommand{\OO}{\mbox{\boldmath $\Omega$} {}}

\newcommand{\SSSS}{\mbox{\boldmath ${\sf S}$} {}}

\newcommand{\MMMM}{\mbox{\boldmath ${\sf M}$} {}}

\newcommand{\ii}{{\rm i}}

\newcommand{\DD}{{\rm D} {}}
\newcommand{\dd}{{\rm d} {}}
\newcommand{\const}{{\rm const}  {}}

\def\half{{\textstyle{1\over2}}}

\def\onethird{{\textstyle{1\over3}}}

\def\quarter{{\textstyle{1\over4}}}
\newcommand{\ypasj}[3]{: #1, {PASJ} {#2}, #3}
\newcommand{\yapj}[3]{: #1, {ApJ} {#2}, #3}
\newcommand{\yapjl}[3]{: #1, {ApJ} {#2}, #3}

\newcommand{\yan}[3]{: #1, {AN} {#2}, #3}
\newcommand{\yact}[3]{: #1, {AcA} {#2}, #3}

\newcommand{\yana}[3]{: #1, {A\&A} {#2}, #3}

\newcommand{\ygafd}[3]{: #1, {GApFD} {#2}, #3}

\newcommand{\ypre}[3]{: #1, {PhRvE} {#2}, #3}

\newcommand{\ymn}[3]{: #1, {MNRAS} {#2}, #3}

\newcommand{\spre}[1]{: #1, PhRvE, submitted}

\newcommand{\yjour}[4]{: #1, {#2} {#3}, #4}

\newcommand{\ybook}[3]{: #1, {\it #2} (#3)}
\newcommand{\yproc}[5]{: #1, in #4 (eds.), {\it #3}, #5, p.~#2}

\newcommand{\sjfm}[1]{: #1, {JFM} (submitted)}

\newcommand{\papj}[1]{: #1, {ApJ} (in press)}

\title{Turbulence and its parameterization in accretion discs}
\author{Axel Brandenburg}
\institute{
NORDITA, Blegdamsvej 17, DK-2100 Copenhagen \O, Denmark\\
}

\date{Received 8 September 2005; accepted 22 September 2005;
published online 20 October 2005}
\begin{document}

\abstract{
Accretion disc turbulence is investigated in the framework of the
shearing box approximation.
The turbulence is either driven by the magneto-rotational instability or,
in the non-magnetic case, by an explicit and artificial forcing term
in the momentum equation.
Unlike the magnetic case, where most of the dissipation occurs in the
disc corona, in the forced hydrodynamic case most of the dissipation
occurs near the midplane.
In the hydrodynamic case evidence is presented for the stochastic
excitation of epicycles.
When the vertical and radial epicyclic frequencies are different
(modeling the properties around rotating black holes), the beat frequency
between these two frequencies appear to show up as a peak in the temporal
power spectrum in some cases.
Finally, the full turbulent resistivity tensor is determined and
it is found that, if the turbulence is driven by a forcing term,
the signs of its off-diagonal components are such that this effect
would not be capable of dynamo action by the shear--current effect.
}

\maketitle

\section{Introduction}

It is now generally accepted that angular momentum transport in accretion
discs is accomplished by hydromagnetic turbulence that is produced by the
magneto-rotational instability (MRI, also known as Balbus-Hawley instability);
see Balbus \& Hawley (1991, 1998).
The significance of this mechanism for accretion discs has been established
using local shearing box simulations (Hawley et al.\ 1995, 1996,
Matsumoto \& Tajima 1995, Brandenburg et al.\ 1995, 1996a, Stone et al.\ 1996),
as well as global simulations (Hawley 2000, Arlt \& R\"udiger 2001,
De Villiers \& Hawley 2003).
For many purposes one would like to parameterize the turbulence in terms
of a turbulent viscosity.
The ultimate goal of such an approach is to be able to capture the
relevant pieces of turbulence physics in two-dimensional axisymmetric
and one-dimensional vertically integrated models of accretion discs.
Even if this turns out not to be possible, parameterized models are
still extremely useful for illuminating otherwise unrecognized mechanisms
that might only be directly identifiable using a targeted approach.

One aspect that we do not understand very well right now is to what extent
MRI-driven turbulence is similar to ordinary (e.g.\ forced) turbulence.
This question is relevant because one is tempted to apply some well-known
turbulence concepts quite loosely also to MRI-driven turbulence without
distinguishing between the different forms of driving.
In order to address this question we consider here the case of forced
turbulence and compare with what has been found for MRI-driven turbulence.
In addition to the forced simulations we also present solutions of MRI-driven
turbulence that have not previously been published.
No net magnetic flux is imposed, so we are able
to have a self-sustained mechanism.
Most previous approaches used numerical viscosity and resistivity.
With the regular laplacian diffusion operator (proportional to $\nabla^2$)
self-sustained turbulence is only possible at large resolution (more than
$128^3$ meshpoints; see Brandenburg et al.\ 2004).
This becomes prohibitively expensive if we need to achieve sufficient
statistics and long enough runs.
Therefore, we adopt hyperdiffusion for these runs (here proportional to
$\nabla^6$), analogous to what was done also in Brandenburg et al.\ (1995).
All forced simulations are however done with regular laplacian diffusion.

One of the goals of this paper is to reconsider the numerical
determination of turbulent viscosity and resistivity in local simulations
of accretion flows.
This is motivated by recent advances in the case of non-shearing and
non-rotating flows.
Particular attention is paid to the tensorial nature of the turbulent
resistivity tensor and the differences between forced and MRI-driven
turbulence.
For the turbulent viscosity the tensorial nature is important for
modeling warps in accretion discs (Torkelsson et al.\ 2000).
This has also motivated the study of the tensorial nature of turbulent
passive scalar diffusion (Carballido et al.\ 2005, Johansen \& Klahr 2005).

In the following we use overbars to denote spatial averages over one
or two coordinate directions, e.g.\ azimuthal averages and occasionally
also vertical averages.
Angular brackets without subscript are used for volume averages, while
angular brackets with subscript $t$ denote time averages.

\section{Stress and strain}

In any parameterized model of a turbulent flow one is interested in the
Reynolds stress $\overline{\rho u_iu_j}$ and, if magnetic fields are
present, in the Maxwell stress $\half\delta_{ij}\overline{\bb^2}
-\overline{b_ib_j}$.
Here, the magnetic field is measured in units where the vacuum permeability
is unity, and lower case characters $\uu$ and $\bb$ denote deviations from
the mean flow $\meanUU$ and the mean field $\meanBB$, so $\meanUU=\UU+\uu$
and $\meanBB=\BB+\bb$ are the full velocity and magnetic fields.
So the full turbulent stress from the small scales (SS) is given by
\EQ
\overline{\Pi}_{ij}^{\rm(SS)}=\overline{\rho u_iu_j}
+\half\delta_{ij}\overline{\bb^2}-\overline{b_ib_j}.
\EN
Analogously, one can define the total stress from the large scale (LS) fields,
\EQ
\overline{\Pi}_{ij}^{\rm(LS)}=\rho\meanU_i\meanU_j
+\half\delta_{ij}\meanBB^2-\meanB_i\meanB_j.
\EN
In the steady state,
the value of the turbulent mass accretion rate, for example,
follows from the constancy of the angular momentum flux, i.e.\
\EQ
\int_0^{2\pi}\varpi\,\dd\phi\int_{-h}^h\dd z\;\left(
\overline{\Pi}_{\varpi\phi}^{\rm(LS)}+\overline{\Pi}_{\varpi\phi}^{\rm(SS)}
\right)=C\equiv\const,
\label{constAMflux}
\EN
where we have neglected the microscopic viscosity, because it is
very small in discs (although not necessarily in simulations!).
Here, cylindrical polar coordinates, $(\varpi,\phi,z)$, have been employed,
and $h$ denotes the disc height (e.g.\ its gaussian scale height).
In \Eq{constAMflux} we can isolate the mass accretion rate,
$\dot{M}=-\int\varpi\,\dd\phi\int\dd z\rho\meanU_\varpi$.
Replacing furthermore the integration by a multiplication with
$4\pi\varpi h$, we find
\EQ
\dot{M}={4\pi h\over \varpi\meanU_\phi}\left(
\varpi\overline{\Pi}_{\varpi\phi}^{\rm(LS)}
-\varpi\meanB_\varpi\meanB_\phi-C\right).
\EN
Remarkably, for calculating the accretion rate it suffices to know
only the value of the stress, not its functional dependence on the mean
flow properties.
The same is true of the heating rate, which is given by
(e.g., Balbus \& Papaloizou 1999, Balbus 2004)
\EQ
\dot{E}=4\pi h\,
\left(\varpi{\partial\meanO\over\partial\varpi}\right)
\left(\overline{\Pi}_{\varpi\phi}^{\rm(LS)}
+\overline{\Pi}_{\varpi\phi}^{\rm(SS)}\right),
\EN
where $\meanO=\meanU_\phi/\varpi$ has been introduced, and only the
shear from the differential rotation has been taken into account.

However, for many (if not all) other purposes it is necessary to know
also the functional dependence of the stress on other quantities,
most notably the shear rate.
In fact, a common proposal is to approximate the mean stress by the
mean rate-of-strain tensor ${\overline{\sf S}}_{ij}$ and to write
\EQ
\overline{\Pi}_{ij}^{\rm(SS)}
\approx2\rho\nu_{\rm t}{\overline{\sf S}}_{ij},
\label{StressStrain}
\EN
where ${\overline{\sf S}}_{ij}=\half(\meanU_{i,j}+\meanU_{j,i})$
is the rate of strain of the mean flow, and $\nu_{\rm t}$ is a
turbulent transport coefficient (turbulent viscosity).\footnote{The
mean flow is here solenoidal, so $\overline{\sf S}_{ij}$ is trace-less.
In the general case considered below an extra term is added to make
sure ${\sf S}_{ij}$ is trace-free.}

The significance of \Eq{StressStrain} is that it provides a closure
for the small scale quantity $\overline{\Pi}_{ij}^{\rm(SS)}$ in terms
of the large scale strain, ${\overline{\sf S}}_{ij}$.
This is, unlike the previous relations in this section, necessarily only
an approximation.
In the equation of angular momentum conservation this term acts as a
diffusion term and provides angular momentum transport in the direction
of decreasing angular velocity, i.e.\ outward.

\begin{figure}[t!]\begin{center}
\includegraphics[width=.9\columnwidth]{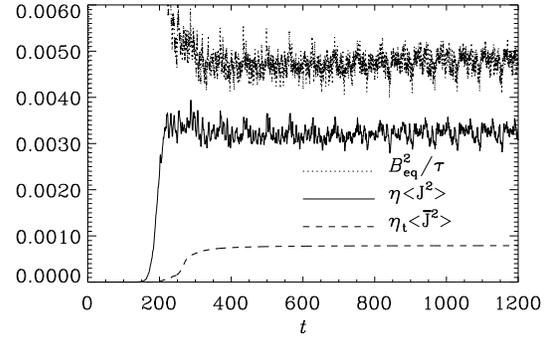}
\end{center}\caption[]{
Joule dissipation for Run~3 of Brandenburg (2001, solid line),
compared with the Joule
dissipation estimated for a corresponding mean-field model (dashed
line). An estimate for the rate of total energy dissipation, $B_{\rm 
eq}^2/\tau$, is also given.
[Adapted from Brandenburg (2003).]
}\label{Fpjoule}\end{figure}

\begin{figure*}[t!]\begin{center}
\includegraphics[width=.75\textwidth]{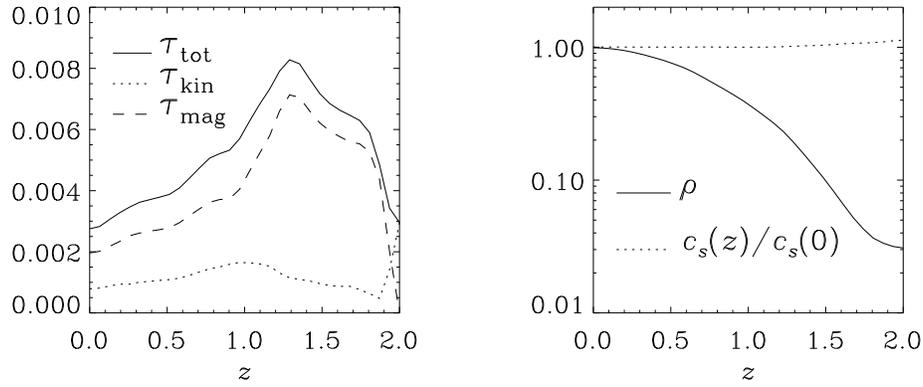}
\end{center}\caption[]{
Dependence of the stress component $\overline{\Pi}_{\varpi\phi}^{\rm(SS)}$
(here denoted by $\tau_{xy}$), separately for the kinetic and magnetic
contributions, together with the sum of the two denoted by total (left)
as well as the vertical dependence of density and sound speed (right).
Note that $\tau_{xy}$ is neither proportional to the density $\rho$ nor
the sound speed $c_{\rm s}$.
[Adapted from Brandenburg et al.\ (1996b).]
}\label{palpz}\end{figure*}

As we discussed above, the stress also contributes to the rate
of heating of the disc.
The horizontally averaged rate of viscous heating (per unit volume)
is given by
\EQ
Q_{\rm visc}=2\rho\nu\overline{\SSSS^2}
\quad\mbox{(actual heating rate)},
\EN
where $\SSSS$ is the actual rate of strain matrix, is assumed to be
approximated by
\EQ
Q_{\rm visc}\approx2\rho\nu_{\rm t}\overline{\SSSS}^2
\quad\mbox{(parameterized heating rate)},
\label{QviscParam}
\EN
where $\overline{\SSSS}$ is the rate of strain of the mean flow.
[In accretion discs, where keplerian shear gives the largest contribution
to $\overline{\SSSS}^2$, \Eq{QviscParam} yields the familiar expression
$Q_{\rm visc}\approx\rho\nu_{\rm t}({3\over2}\meanO)^2$; see
Frank et al.\ (1992).]
Of course, $\overline{\SSSS^2}\gg\overline{\SSSS}^2$ and
$\nu_{\rm t}\gg\nu$, but whether $\nu_{\rm t}\overline{\SSSS}^2$
is actually the same as $\nu\overline{\SSSS^2}$ is not obvious.

Using the first order smoothing approximation, which is commonly
used in mean field dynamo theory, R\"udiger (1987) found that
$\nu\overline{\SSSS^2}$ is actually about 3 times larger than
what is expected from $\nu_{\rm t}\overline{\SSSS}^2$.
It is still unclear whether this is actually true, or whether it is
an artifact of the first order smoothing approximation and that
the two expressions should really be the same.

In quite a different context of hydromagnetic turbulence, where a
helical large scale mean field is generated (so that the mean current
density is well defined), an equivalent conclusion was reached
for the actual and parameterized {\it resistive} heating rates, i.e.\
\EQ
\eta\overline{\JJ^2}\approx3.7\times\eta_{\rm t}\meanJJ^2,
\label{QviscRel}
\EN
see Brandenburg (2003) and \Fig{Fpjoule} where we also compare with an
estimate for the rate of total energy dissipation, $B_{\rm eq}^2/\tau$,
where $\tau$ is the turnover time.
Here we have used for $\eta_{\rm t}$ the value estimated from the
self-consistently determined empirical quenching formula of
Brandenburg (2001) for his Run~3.
The factor $3.7$ in \Eq{QviscRel} is similar to what has been found
for the turbulent viscosity (R\"udiger 1987).
This suggests that his result obtained within the framework of the
first order smoothing approximation is at least not in conflict with
the simulations.

We now turn to another aspect of viscous heating.
In MRI-driven turbulence it was found that the stress does not decrease
with height away from the midplane, as suggested by \Eq{StressStrain},
even though the product of rate of strain and density does
decrease because of decreasing density.
This was originally demonstrated only for nearly isothermal
discs (Brandenburg et al.\ 1996b), see \Fig{palpz}, but this has
now also been shown for radiating discs (Turner 2004).

\begin{figure}[t!]\begin{center}
\includegraphics[width=.95\columnwidth]{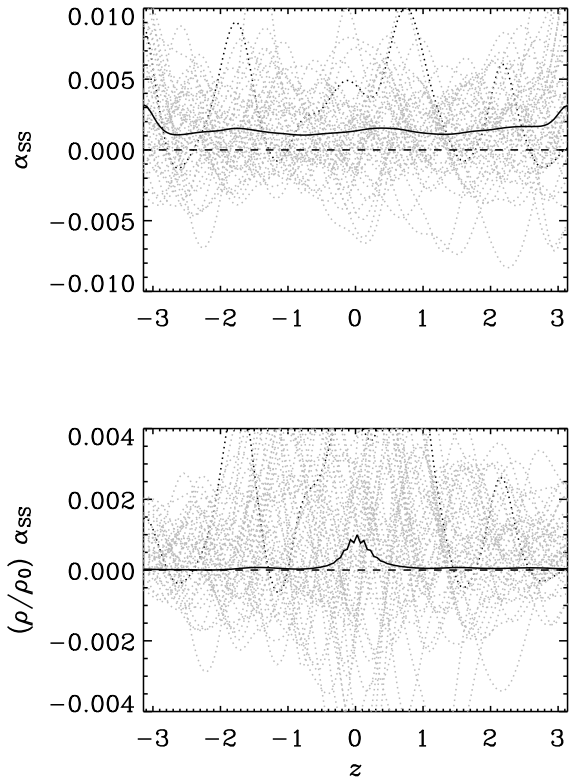}
\end{center}\caption[]{
Vertical dependence of $\alpha_{\rm SS}$ for hydrodynamic forced shear flow
turbulence (upper panel).
The solid curves give the time average while a few individual times are
plotted in gray.
Note that the temporal average of $\alpha_{\rm SS}$ is nearly constant in $z$,
although for individual times there can be significant fluctuations.
By comparison, the density-weighted viscosity parameter decreases with height
(lower panel), showing that in forced non-magnetic turbulence the stress
is indeed proportional to the density.
}\label{ppstress}\end{figure}

Indeed, it was found that $\rho$ should rather be replaced by the
vertically averaged density.
A sensible parameterization seemed therefore only possible for the
vertically integrated stress, i.e.\ (Brandenburg et al.\ 1996b)
\EQ
\int_{-h}^h\dd z\overline{\Pi}_{\varpi\phi}^{\rm(SS)}
\approx\alpha_{\rm SS}\int_{-h}^h\dd z\,\rho c_{\rm s}^2,
\EN
where $\alpha_{\rm SS}$ is the dimensionless Shakura-Sunyaev (1973)
viscosity coefficient and $c_{\rm s}$ is the sound speed.
The reason for this is quite clearly related to the fact that the
stress is of magnetic origin, and that most of the dissipation happens
when the density is low, i.e.\ when the ohmic heating rate per unit mass,
$\JJ^2/(\sigma\rho)$, is large (here, $\JJ$ is the current density, and
$\sigma$ the electric conductivity).
However, it would be surprising if the absence of a scaling of the stress
with the density were a general property.

In order to shed some light on this question we now consider a
hydrodynamic shearing box model where the turbulence is driven by an
explicit body force that is delta-correlated in time and monochromatic
in space with a wavenumber of $k_{\rm f}\approx3$.
(Details of this simulation are given in \Sec{ShearingBox}.)
The result is shown in \Fig{ppstress}, where we have expressed the result
in terms of a vertically dependent Shakura-Sunyaev viscosity coefficient
$\alpha_{\rm SS}(z)$, so
\EQ
\nu_{\rm t}(z)=\alpha_{\rm SS}(z)\,c_{\rm s}h.
\EN
Note that in the present case of purely hydrodynamic forced turbulence
$\alpha_{\rm SS}(z)$ is nearly independent of height, even though $\rho$
varies by at least an order of magnitude, as can be seen from the second
panel of \Fig{ppstress}, where we show the product
$\rho(z)\alpha_{\rm SS}(z)$, normalized by the initial density in the
midplane, $\rho_0$.

In the following section we reiterate the standard shearing box equation
that have been used to obtain this result.

\section{Shearing box equations}
\label{ShearingBox}

We consider here a domain of size $L_x\times L_y\times L_z$,
where $L_x=L_y=2\pi$ and $L_z=4$, although other choices
are possible and have been considered in the earlier work.
We solve the equations for the departure from the keplerian shear flow
for an isothermal equation of state (so the pressure is given by
$p=\rho c_{\rm s}^2$).
The resulting equations can be written in the form
\EQA
{\DD u_i\over\DD t}=-u_j{\partial\meanU_i\over\partial x_j}
+\left({\JJ\times\BB\over\rho}-c_{\rm s}^2\nab\ln\rho-2\OO\times\uu\right)_i
\nonumber \\
-\zeta^2 z_i+f_i
+\nu\left(\nabla^2 u_i+\onethird{\partial\over\partial x_i}\nab\!\cdot\!\uu
+2{\sf S}_{ij}{\partial\ln\rho\over\partial x_j}\right)\!,
\ENA
\EQ
{\DD A_i\over\DD t}=-A_j\left({\partial\meanU_j\over\partial x_i}
+{\partial u_j\over\partial x_i}\right)+\eta\nabla^2A_i\,,
\EN
\EQ
{\DD\ln\rho\over\DD t}=-\nab\cdot\uu,
\EN
where $\DD/\DD t=\partial/\partial t+(\meanUU+\uu)\cdot\nab\,$ is
the advective derivative based on the full flow field that includes
both the shear flow and the deviations from it.
We solve the equations for the departure, $\uu$, from the purely linear
shear flow.
Except for the advection operator, only derivatives of $\meanUU$ enter.
This is an important property of the shearing sheet approximation that
is critical for being able to use shearing-periodic boundary conditions
in the $x$ direction.
We have used the gauge
\EQ
\phi=\eta\nab\cdot\AAA-(\meanUU+\uu)\cdot\AAA,
\quad\mbox{(assuming $\eta=\const$)}.
\EN
for the electrostatic potential.
The shear flow is given by $\meanUU(x)=(0,-q\Omega x,0)$,
where $q=3/2$ for a purely keplerian shear flow, and
\begin{equation}
{\sf S}_{ij}=
\frac{1}{2}\left({\partial u_i\over\partial x_j}
+ {\partial u_j\over\partial x_i}\right)
-\frac{1}{3}\delta_{ij}\nab\cdot\uu
\end{equation}
is the traceless rate of strain matrix.
The Coriolis force is added to take into account that the
shearing box is spinning about the central star at
angular velocity $\Omega$.
This together with shear can lead to radial epicyclic
oscillations with frequency
\EQ
\kappa=\sqrt{2(2-q)}\,\Omega.
\EN
Note that $\kappa=\Omega$ for a keplerian disc with $q=3/2$.
The term $-\zeta^2\zz$ characterizes the vertical stratification, where
$\zeta$ is the vertical epicyclic frequency.
Again, in a keplerian disc we have $\zeta=\Omega$, but here we treat
$\zeta$ as an independent parameter in order to assess the effects
of different radial and vertical epicyclic frequencies in non-newtonian
discs with different radial and vertical epicyclic frequencies
(Abramowicz et al.\ 2003a,b, Kato 2004,
Klu\'zniak et al.\ 2004, Lee et al.\ 2004).
For most of the cases considered below we choose $\Omega=\kappa=0.4$ and
$\zeta=0.6$.
The sound speed is taken to be $c_{\rm s}=1$.
(With these parameters the magnitude of the shear flow is
${3\over2}\meanO(L_x/2)\approx1.9$, so it is weakly supersonic.)
In the nonmagnetic cases ($\AAA=0$)
we allow for the possibility of an extra forcing term $\ff$ in
order to study the case of non-magnetic (non-MRI) turbulence.
The forcing function is identical to that used by Brandenburg (2001)
for forced turbulent simulations exhibiting dynamo action.
The amplitude of the forcing function is denoted by $f_0$ which
is here chosen to be 0.01, unless it put to zero.
The details of this forcing function are summarized in Appendix~A.

The boundary conditions adopted in the vertical direction on
$z=\pm L_z/2$ are
\EQ
u_{x,z}=u_{y,z}=u_z=A_x=A_y=A_{z,z}=0,
\EN
which corresponds to a perfect conductor no-slip condition.
Here, commas denote partial differentiation.

The results presented here have been obtained with the
\textsc{Pencil Code},\footnote{%
\url{http://www.nordita.dk/software/pencil-code}}
which is a public domain code for solving the compressible hydromagnetic
(and other) equations on massively parallel distributed memory
architectures such as Beowulf clusters.
It employs a high-order finite-difference scheme (sixth order in space
and third order in time), which is ideal for all types of turbulence
simulations.

The present simulations are mainly exploratory in nature, so we only
use small to moderate resolution between $32^3$ and $128^3$ meshpoints.
Restrictions in resolution are also imposed by the need to run for
long enough times in order to overcome transients.

\section{Growth and decay of epicycles}

One way of estimating the effective viscosity of a system is to determine
the decay rate of a velocity perturbation that has initially a sinusoidal
variation in one spatial direction.
As an example one may consider an initial mean velocity perturbation
of the form
\EQ
\uu(\xx,0)=\meanUU_0(0)\sin\kk_i\cdot\xx
\EN
for different directions $i$ of the wavevector $\kk_i$.
The time dependence of the amplitude $\meanUU_0(t)$ can be determined
by projecting on the original velocity, i.e.\
\EQ
\meanUU_0(t)=2\bra{\meanUU(\xx,t)\sin\kk_i\cdot\xx},
\EN
where the angular brackets denote volume averaging.
The effective viscosity is connected with the decay rate $\lambda$
via $\lambda=\nu_{\rm t}\kk_i$, and $\lambda$ is measured as
\EQ
\lambda=-\bra{\dd\ln|\meanUU_0(t)|/\dd t}_t,
\EN
where $\bra{.}_t$ denotes a time average over a suitable time
interval when the decay is exponential.
In forced turbulence it was found that
$\nu_{\rm t}\approx(0.8\ldots0.9)\times u_{\rm rms}/k_{\rm f}$
(Yousef et al.\ 2003).

\begin{figure}[t!]\begin{center}
\includegraphics[width=\columnwidth]{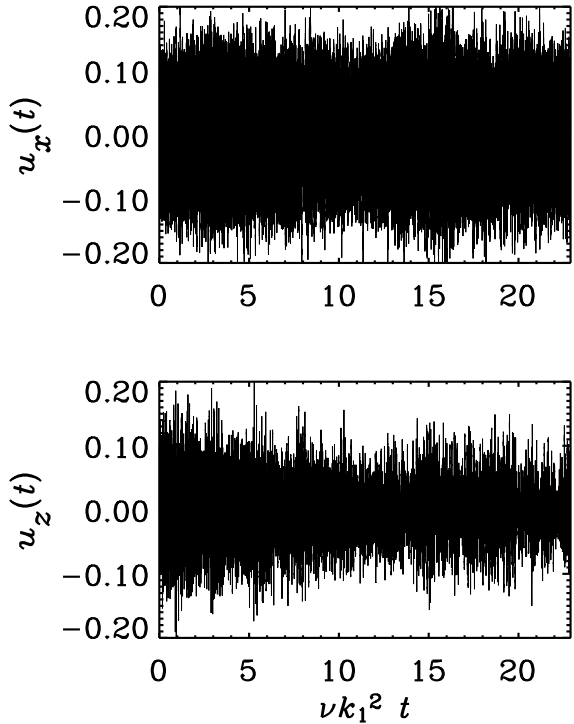}
\end{center}\caption[]{
Time dependence of the vertical and radial velocity components
at one point.
Note that time is represented in viscous units, where $2\pi/k_1$
is the vertical extent of the computational domain.
Evidently, the decay of the vertical epicyclic oscillations in $u_z$
is very small (lower panel), and there is no decay of $u_x$ (upper panel).
}\label{pt}\end{figure}

\begin{figure}[t!]\begin{center}
\includegraphics[width=\columnwidth]{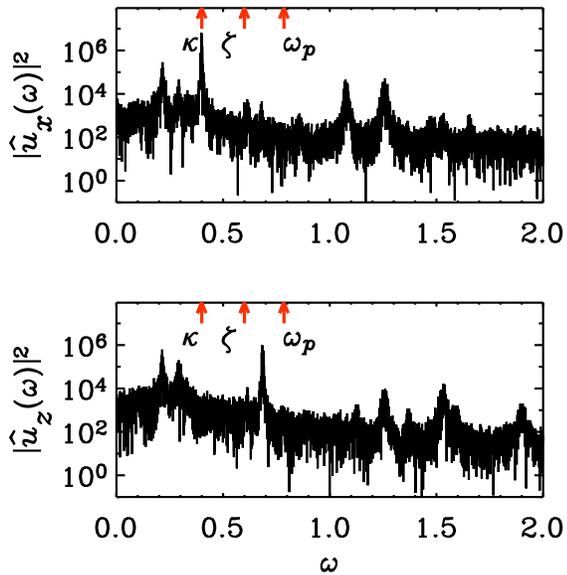}
\end{center}\caption[]{
Power spectra of $u_x$ and $u_z$.
The orbit of the initial state is perturbed, using $\uu=(0.1, 0, 0.1)$,
such that radial and vertical epicyclic motions are excited with
frequencies $\kappa=\Omega=0.4$ and $\zeta=0.6$
(marked by arrows in the two diagrams).
The frequency of standing sound waves (p-modes) is labeled as $\omega_p$.
The forcing amplitude is $f_0=0.01$.
}\label{pogood}\end{figure}

\begin{figure}[t!]\begin{center}
\includegraphics[width=\columnwidth]{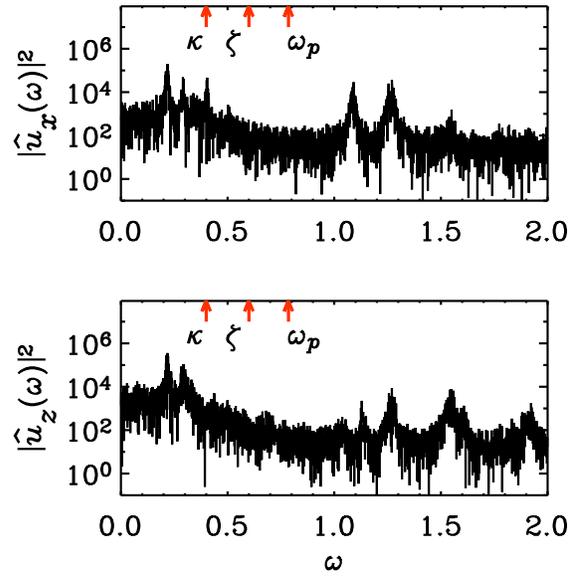}
\end{center}\caption[]{
Same as \Fig{pogood}, but without initial velocity perturbation
corresponding to a perfectly circular orbit orbit.
Note that the lower beat frequency, $\omega_{\rm l}=\zeta-\kappa=0.2$,
appears to be excited in both horizontal and vertical velocity components,
while the upper beat frequency, $\omega_{\rm u}=\zeta+\kappa=1$, appears
only in the radial velocity component.
}\label{pogood_randec64d_ampl4}\end{figure}

However, it turned out that in the present case the initial perturbations
drive epicyclic oscillations that, once excited, do not easily decay.
In \Fig{pt} we show the evolution of the radial and vertical velocity
components at one point ($u_x$ and $u_z$, respectively) in viscous units.
Here the initial velocity perturbation was chosen to be $\uu=(0.1, 0, 0.1)$.
The initial $x$ velocity is compatible with the shearing periodic boundary
conditions, but a uniform vertical velocity is not.
This turns out to be not a problem, because viscosity is able to damp
the initially produced sharp gradients that occur near the boundary to
satisfy $u_z=0$.

Looking at \Fig{pt} it is clear that there is hardly any decay.
On the other hand, if no perturbation is applied initially,
some oscillations are still being generated by the turbulence,
as can be seen from the temporal Fourier spectra of the radial
and vertical velocity components at one point in the domain,
$|\hat{u}_x(\omega)|^2$ and $|\hat{u}_z(\omega)|^2$, respectively.
These are shown in \Fig{pogood} and \Fig{pogood_randec64d_ampl4}
for the cases with and without initial perturbations that
would initialize epicyclic oscillations.

Indeed, in \Fig{pogood} one sees quite clearly the radial epicyclic
frequency $\kappa=0.4$ in the radial velocity component and the vertical
epicyclic frequency $\zeta=0.6$ in the vertical velocity component.
However, the latter seems to be shifted toward a somewhat higher frequency
(about 0.7), which might be due to the interaction with the vertical
p-mode frequency, $\omega_p=c_{\rm s}\pi/L_z\approx0.78$.

It turns out that even when the oscillations are not present initially,
some discrete frequencies are still being excited that lie near the lower
and upper beat frequencies ($\omega_{\rm l}=\zeta-\kappa=0.2$ and
$\omega_{\rm u}=\zeta+\kappa=1$; see \Fig{pogood_randec64d_ampl4}.
However, the match is only approximate, suggesting that the cause of
the additional peaks in the spectra might not necessarily be related
to beat phenomena.

It is plausible that epicyclic oscillations in discs can be
excited stochastically.
This is analogous to the excitations of p-modes in the sun and in
stars (Goldreich \& Kumar 1990).
Similar arguments may also be applied to discs in order to explain the
quasiperiodic oscillations (QPOs) in terms of resonances
between different vertical and horizontal epicyclic frequencies
(Abramowicz et al.\ 2003a,b, Kato 2004,
Klu\'zniak et al.\ 2004, Lee et al.\ 2004).

\begin{figure}[t!]\begin{center}
\includegraphics[width=\columnwidth]{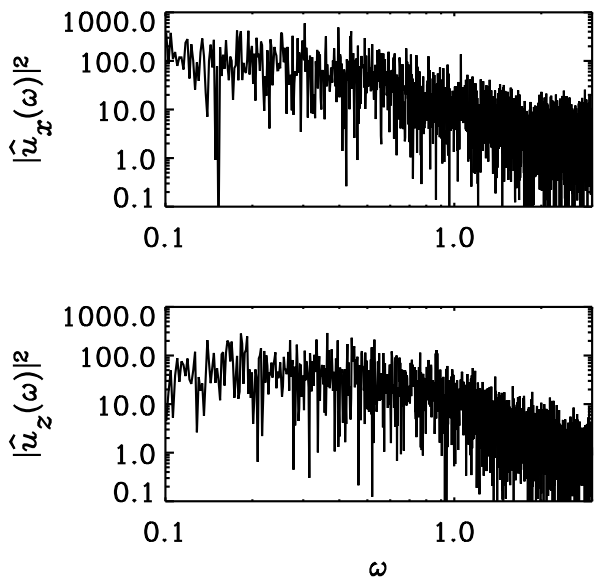}
\end{center}\caption[]{
Same as \Fig{pogood}, but with magnetic field and no forcing,
and in double-logarithmic representation.
Thus, turbulence is generated by the MRI.
As in \Fig{pogood}, an initial perturbation of amplitude 0.1 has
been applied.
Note that for low frequencies the spectrum is nearly flat and lacks
any discrete frequencies.
}\label{pogood_mri32d_tst2}\end{figure}

Next we compare with a case where magnetic fields are included
so that turbulence can be produced as a result of the MRI.
No forcing function is therefore applied in the momentum equation.
It turns out that in this case the power spectrum is more nearly flat
and does not show discrete frequencies as in the purely hydrodynamic
cases with an explicit forcing function; see \Fig{pogood_mri32d_tst2}.

\begin{figure}[t!]\begin{center}
\includegraphics[width=\columnwidth]{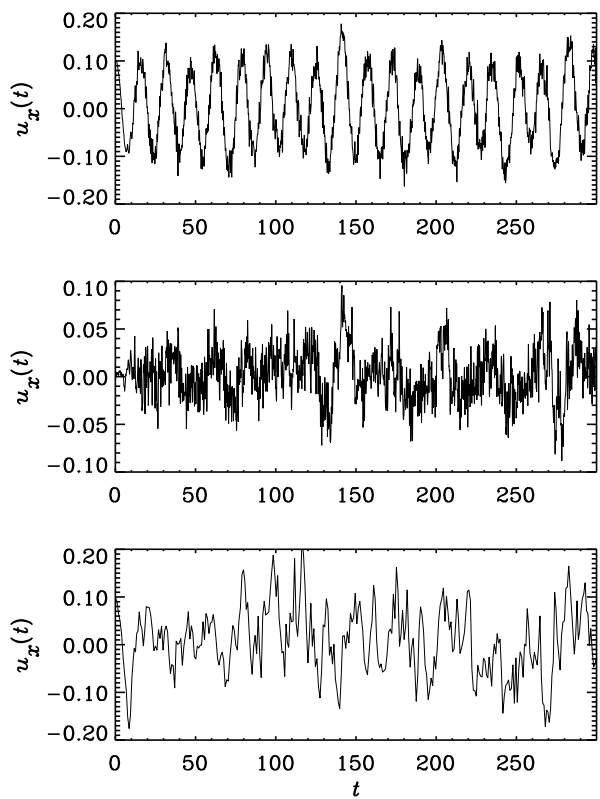}
\end{center}\caption[]{
Comparison of the time series for the runs shown in
\FFigs{pogood}{pogood_mri32d_tst2}.
Only the beginning of the time series are shown.
Note that in the last panel the MRI driven turbulence
quickly wipes out the initial perturbation and no
epicyclic oscillations are visible.
}\label{p300}\end{figure}

The absence of discrete frequencies in the velocity
power spectrum of MRI-driven turbulence is surprising.
In order to inspect further what happens after the initial kick
that was imposed via the initial condition, we plot
in \Fig{p300} the time evolution for all three runs
during the first 300 time units.
(The total duration of these runs is around 10,000 time units.)
It is clear from the figure that in the case of forced turbulence
the amplitude of the epicycles remains dominant.
In the case of MRI-driven turbulence, the epicycles are essentially
swamped by the comparatively strong level of turbulence.
By comparison, in the nonmagnetic case of forced turbulence
some systematic oscillation pattern is visible even when there
is no initial perturbation giving rise to epicycles.

\begin{figure}[t!]\begin{center}
\includegraphics[width=\columnwidth]{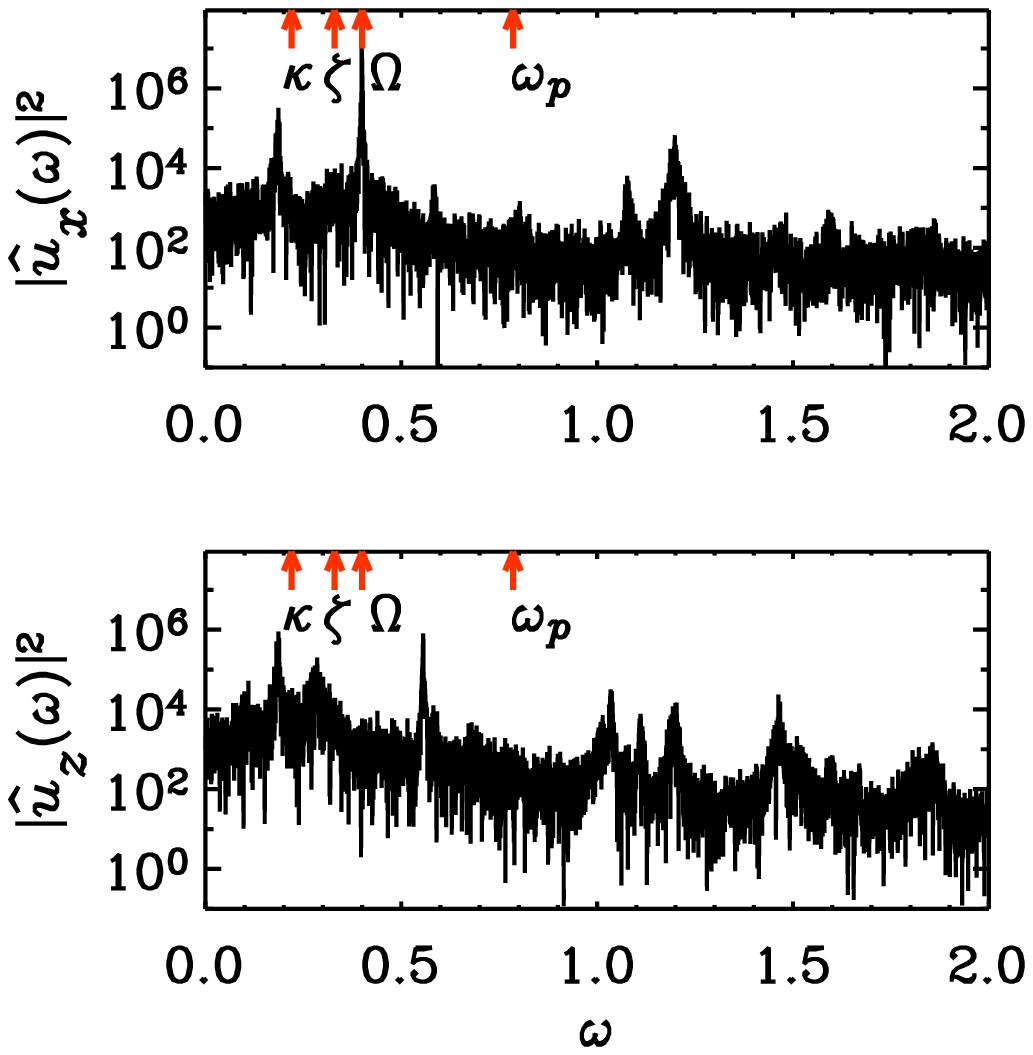}
\end{center}\caption[]{
Same as \Fig{pogood}, but
$\kappa=0.22$, $\zeta=0.33\approx{3\over2}\kappa$,
and $\Omega=0.4$.
}\label{pogood_randec64d_q185}\end{figure}

Finally, we present a case where the shear parameter $q$ is
chosen to be 1.85, so the radial epicyclic frequency is
now different from the rotation frequency, i.e.\
$\kappa=0.22$, while $\Omega=0.4$.
The vertical epicyclic frequency is chosen to be
$\zeta=0.33\approx{3\over2}\kappa$.
It turns out that there is still a pronounced peak at 0.4 ($=\Omega$),
while some of the earlier detected frequencies (0.2 and 1.0), which
are still present, lost their originally anticipated interpretation;
see \Fig{pogood_randec64d_q185}.
The vertical velocity has no longer a peak at $\zeta=0.33$, but instead
at a higher frequency of about $0.5$.
However, it appears still possible to interpret this as some
interplay between $\zeta=0.33$ and $\omega_p=0.78$.
The frequency $0.2$ is still visible in the spectra, but with the
new values of $\kappa$ and $\zeta$ it is no longer possible to
explain this as the lower beat frequency between the two.
Although this result is perhaps somewhat disappointing in terms
of interpreting QPOs, these experiments highlight the
general usefulness of using the shearing sheet approximation which
allows these kinds of experiments to be carried out without difficulties.

The above discussion is interesting in its own right and
deserves certainly more attention.
However, this method is obviously not well suited for determining turbulent
transport coefficients.
Therefore we now describe another method that has recently
been developed in the hydromagnetic context for determining
dynamo parameters.

\section{Tensorial turbulent resistivity}
\label{Tensorial}

The theory of turbulent resistivity is in many ways more
developed than the theory of turbulent viscosity.
In dynamo theory it has recently become possible to determine
quite accurately not only the turbulent resistivity, but also
its full tensorial form and other components that can be
non-dissipative and hence important for dynamo action.
While in the hydrodynamic case one is interested in the correlation
$\overline{u_iu_j}$, one is here interested in the correlation
$\overline{u_jb_j}$, or more specifically in the electromotive force
$\meanemf_i=\epsilon_{ijk}\overline{u_jb_j}$.
Assuming that the mean field is spatially smooth (which may not be the case
in practice) one can truncate the expression for $\meanemf_i$ in terms
of $\meanB_j$ and its derivatives after the first derivative, so one has
\EQ
\meanemf_i=\alpha_{ij}\meanB_j+\eta_{ijk}\meanB_{j,k}.
\label{emfExpression}
\EN
The components of $\alpha_{ij}$ tensor are usually quite easily determined
from simulations by imposing a uniform magnetic field $\meanB_j$
and measuring the resulting electromotive force $\meanemf_i$, so that
$\alpha_{ij}=\meanemf_i/\meanB_j$ is obtained straightforwardly.
The reason this works is because for a uniform field all derivatives
of $\meanB_j$ vanish, so there are no higher order terms.
Calculating the components of $\eta_{ijk}$ is usually harder,
especially when the mean field may no longer be smooth and its
derivatives may vanish in places.
A method that has been used for accretion disc turbulence is based
on a fitting procedure of the measured mean field and the mean
electromotive force to \Eq{emfExpression} by calculating moments of
the form $\bra{\meanemf_i\meanB_j}$, $\bra{\meanemf_i\meanB_{j,k}}$,
as well as $\bra{\meanB_i\meanB_j}$ and $\bra{\meanB_i\meanB_{j,k}}$
(Brandenburg \& Sokoloff 2002, Kowal et al.\ 2005).

A general procedure for determining the full $\alpha_{ij}$ and
$\eta_{ijk}$ tensors from a simulation is to calculate the electromotive
force after applying test fields of different directions and with
different gradients (Schrinner et al.\ 2005).
In the following we adopt $xy$ averages, so the resulting mean fields
depend only on $z$ and $t$, and only $\meanB_x$ and $\meanB_y$ are
non-trivial ($\meanB_z=0$ because of the solenoidality $\meanBB$).
Therefore, only the four components of $\alpha_{ij}$ and the four
components of $\eta_{ij3}$ with $i,j=1,2$ are non-trivial.
Here, the numbers $1,2,3$ refer to the cartesian $x,y,z$ components.

In the present case of one-dimensional mean fields it is
advantageous to rewrite \Eq{emfExpression} in the form
\EQ
\meanemf_i=\alpha_{ij}\meanB_j-\tilde{\eta}_{ij}\meanJ_j,
\quad i,j=1,2,
\EN
where $\meanJJ=\nab\times\meanBB$ is the mean current density, and
\EQ 
\tilde{\eta}_{il}=\eta_{ijk}\epsilon_{jkl}
\label{tildeETArel}
\EN
is the resistivity tensor operating only on the mean current density.
In the special case of one-dimensional averages there is no extra
information contained in the symmetric part of the $\meanB_{j,k}$
tensor that is not already contained in the components of $\meanJJ$.
In fact, the four components of $\eta_{ij3}$ map uniquely to those
of $\tilde{\eta}_{il}$ via
\EQ
\pmatrix{
\tilde\eta_{11}& \tilde\eta_{12}\cr
\tilde\eta_{21}& \tilde\eta_{22}}
=\pmatrix{
\eta_{123} & -\eta_{113} \cr
\eta_{223} & -\eta_{213} }.
\EN
This fact was also used in Brandenburg \& Sokoloff (2002).
The diagonal components of $\tilde{\eta}_{ij}$ correspond to
turbulent resistivity, while its off-diagonal components can
be responsible for driving dynamo action [$\OO\times\meanJJ$
and $\meanWW\times\meanJJ$ effects; see, R\"adler (1969) and
Rogachevskii \& Kleeorin (2003, 2004), R\"adler \& Stepanov (2005)].
Conversely, the diagonal components of the $\alpha$ tensor
can be responsible for dynamo action while the off-diagonal
components are responsible for non-regenerative
turbulent pumping effects (Krause \& R\"adler 1980).
It should be noted, however, that for linear shear flows
R\"udiger \& Kitchatinov (2005) find that the signs of the relevant
coefficients of $\tilde{\eta}_{ij}$ are such that dynamo action is not
possible for small magnetic Prandtl numbers.

In summary, in the present case of one-dimensional mean fields,
$\meanBB=\meanBB(z,t)$, there are altogether $4+4$ unknowns.
The idea is to calculate the electromotive force
\EQ
\meanEMF^{(p,q)}=\overline {\uu\times\bb^{(p,q)}}
\label{emftest}
\EN
for the excess magnetic fluctuations, $\bb^{(p,q)}$, that are due
to a given test field $\meanBB^{(p,q)}$, where the labels $p$ and $q$
characterize the test field ($p$ gives its nonvanishing
component and $q=1$ or 2 stands for cosine or sine-like test fields.
The calculation of the electromotive force
requires solving simultaneously a set of equations of the form
\EQ
{\partial\bb^{(p,q)}\over\partial t}=\nab\times\left[
\left(\meanUU+\uu\right)\times\meanBB^{(p,q)}
\right]+\eta\nabla^2\bb^{(p,q)}+\GG
\EN
for each test field $\meanBB^{(p,q)}$.
Here, $\GG=\nab\times[\uu\times\bb^{(p,q)}-\overline{\uu\times\bb^{(p,q)}}]$
is a term nonlinear in the fluctuation.
This term would be ignored in the first order smoothing approximation,
but it can be kept in a simulation if desired.
(In the present considerations it is neglected.)

The four test fields considered in the present problem of
one-dimensional mean fields are
\EQ
\meanBB^{(1,1)}=\pmatrix{\cos k_1 z\cr0\cr0},\quad
\meanBB^{(1,2)}=\pmatrix{\sin k_1 z\cr0\cr0},
\EN
\EQ
\meanBB^{(2,1)}=\pmatrix{0\cr\cos k_1 z\cr0},\quad
\meanBB^{(2,2)}=\pmatrix{0\cr\sin k_1 z\cr0}.
\EN
As an example, consider the $x$ component of $\meanEMF^{(p,q)}$ for
$p=1$ and both values of $q$,
\EQA
\meanemf_1^{(1,1)}=\alpha_{11}\cos k_1 z-\eta_{113}\sin k_1 z,
\\
\meanemf_1^{(1,2)}=\alpha_{11}\sin k_1 z+\eta_{113}\cos k_1 z.
\ENA
For $p=2$, and/or for $i=2$, one obtains a similar pair of
equations with the same arrangement of cosine and sine functions.
So, for each of the four combinations of $i$ and $j$ ($=p$) the set of two
coefficient, $\alpha_{ij}$ and $\eta_{ij3}$, is obtained as
\EQ
\pmatrix{\alpha_{ij}\cr\eta_{ij3}}=\MMMM^{-1}
\pmatrix{\meanemf_i^{(j,1)}\cr\meanemf_i^{(j,2)}},
\EN
where the matrix
\EQ
\MMMM=\pmatrix{
\cos k_1 z & -\sin k_1 z\cr
\sin k_1 z & \cos k_1 z}
\EN
is the same
for each value of $p$ and each of the two components $i=1,2$
of $\meanemf_i^{(p,q)}$.
Finally, $\tilde\eta$ is calculated using \Eq{tildeETArel}.
Note that $\det\MMMM=1$, so the inversion procedure is well behaved
and even trivial.

\begin{figure}[t!]\begin{center}
\includegraphics[width=\columnwidth]{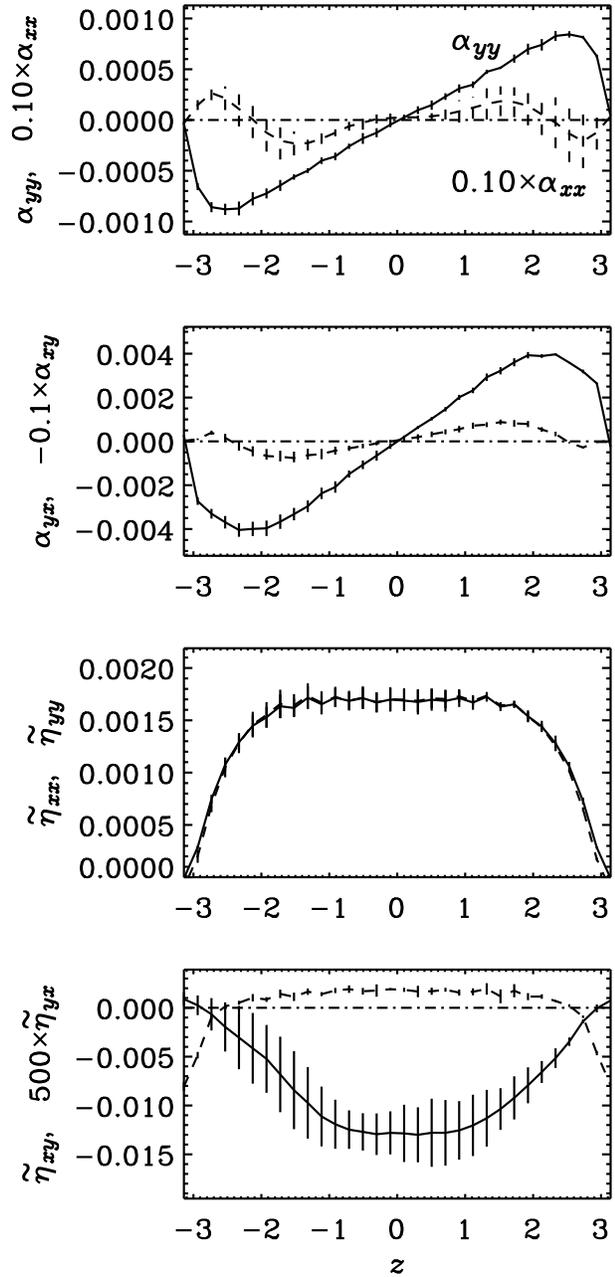}
\end{center}\caption[]{
Magnetic turbulent transport coefficients for forced turbulence.
Solid and dashed lines refer respectively to the first and second
quantity denoted on the corresponding axis.
The error bars have been obtained by calculating the maximum
departure over the three possibilities obtained by considering
only 1/3 of the full time series.
}\label{palp_aver2}\end{figure}

\begin{figure}[t!]\begin{center}
\includegraphics[width=\columnwidth]{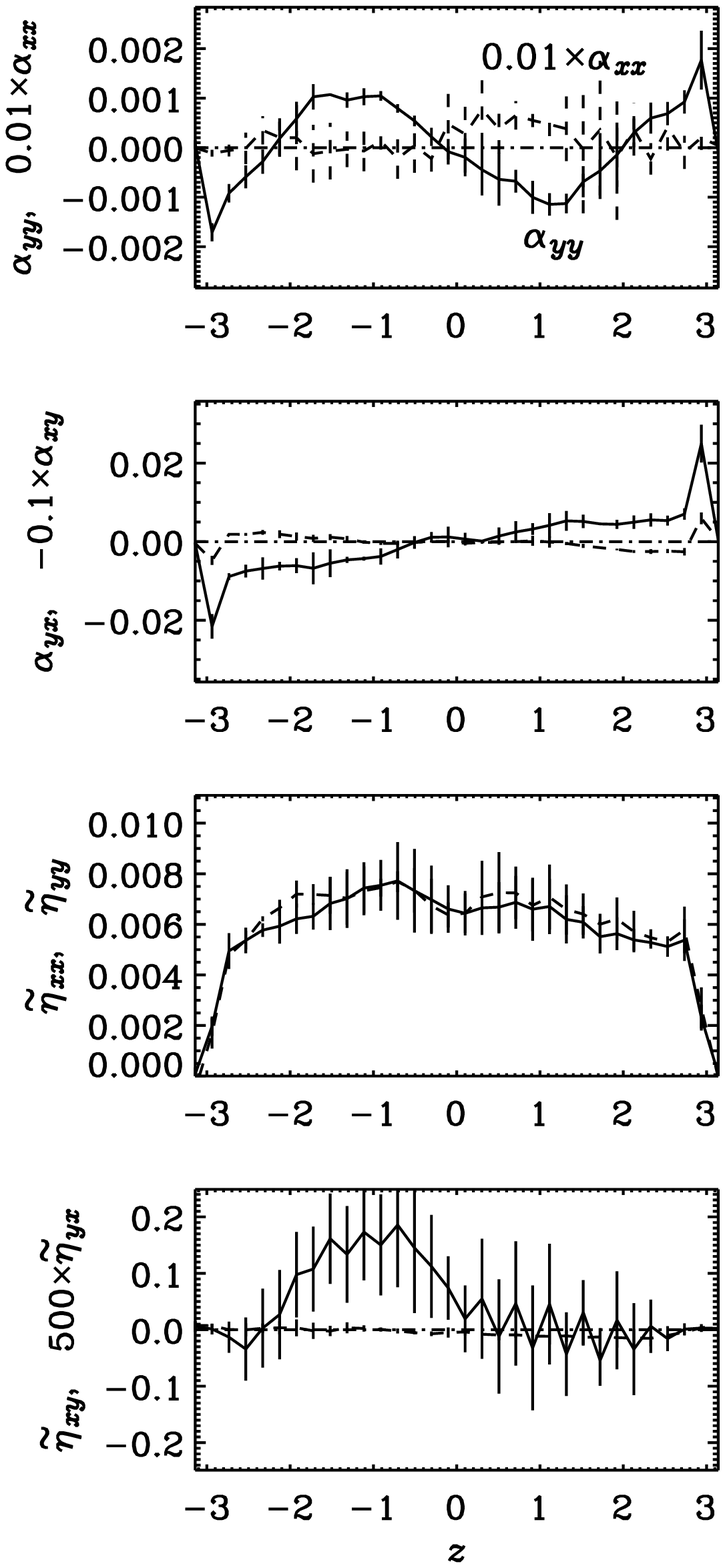}
\end{center}\caption[]{
Same as \Fig{palp_aver2}, but for MRI-driven turbulence.
}\label{palp_aver2_mri32d_tst2}\end{figure}

The test field algorithm described above has been implemented
in the \textsc{Pencil Code}.
The results are shown in \Fig{palp_aver2} for the case of forced
turbulence and in \Fig{palp_aver2_mri32d_tst2} for the case of
MRI-driven turbulence.
We recall that in both cases the effect of stratification is included.

The following main results can be summarized from this analysis.
First, for forced turbulence the $\alpha$ effect (especially the $yy$
component that is relevant for acting on the already strong toroidal
$\meanB_y$ field) is {\it positive} in the northern hemisphere $z>0$
[as expected for cyclonic and anti-cyclonic events; see Parker (1979)
and Krause \& R\"adler (1980)].
However, in the case of MRI-driven turbulence the sign of $\alpha_{yy}$
is reversed, confirming the early results of Brandenburg et al.\ (1995).
This may be explained in terms of strong flux tubes that are buoyant
and therefore rising, but that also exhibit a conver\-ging flow
from the $\BB\cdot\nab\BB$ tension force that points toward to
strongest parts of the tube (Brandenburg 1998).
The same result has been obtained by R\"udiger \& Pipin (2000)
for magnetically driven turbulence using first order smoothing.
Their result is specifically due to the dominance of the current
helicity term and is not connected with the buoyancy term (Blackman 2005).

Second, the off-diagonal components of the $\alpha$ tensor are such
that they correspond to a turbulent pumping effect,
$\meanEMF=...+\ggamma\times\meanBB$, where
$\gamma_z=\half(\alpha_{yx}-\alpha_{xy})$.
Theoretically the pumping velocity is expected to be in the direction
of negative turbulent intensity (Roberts \& Soward 1975).
In the present case of forced turbulence the rms velocity is
approximately independent of height, but the density decreases
outwards, causing therefore an outward turbulent pumping effect
(second panel of \Fig{palp_aver2}).
For MRI-driven turbulence, the density also deceases outward, but
the rms velocity increases as the density decreases such that
$\rho\overline{\uu^2}$ is approximately constant.
This may be the reason for not having much of a vertical turbulent
pumping effect in MRI-driven turbulence.

Third, the two diagonal components of $\tilde{\eta}_{ij}$ are in both
cases positive (i.e.\ indeed diffusive, which is non-trivial!) and
the two components, $\tilde{\eta}_{xx}$ and $\tilde{\eta}_{yy}$,
are almost equally big.
[We recall that in Brandenburg \& Sokoloff (2002) it was found that
$\tilde{\eta}_{yy}$ (responsible for diffusion of $\meanB_x$)
was very small, but this result was perhaps not accurate.]

Fourth, the signs of the off-diagonal components of $\tilde{\eta}_{ij}$
are here such that they would not correspond to a dynamo effect
of the form $\meanEMF=...+\ddelta\times\meanJJ$, where
$\delta_z=\half(\tilde\eta_{xy}-\tilde\eta_{yx})$.
Growing solutions require that the product of shear
(here $S=-{3\over2}\Omega$) and $\delta_z$ is negative
(Brandenburg \& Subramanian 2005a,b).
However, since $\delta_z<0$ for forced turbulence (\Fig{palp_aver2}),
only decaying solutions are possible.
This is in agreement with predictions by R\"udiger \& Kitchatinov (2005).
For MRI-driven turbulence $\delta_z$ is within error bars
either compatible with zero or positive in a few places.
Thus, the $\meanWW\times\meanJJ$ effect is perhaps possible here.
However, for the full $2\times2$ tensor, $\tilde\eta_{ij}$, it is
important to consider its tensorial nature.
It turns out that a necessary condition for dynamo action is
\EQ
\tilde\eta_{yx}k_z^2(\tilde\eta_{xy}k_z^2\!+\!S)>0
\quad\mbox{(for dynamo action)},
\EN
so the sign of $\tilde\eta_{yx}$ is now crucial; see Appendix B for details.
Even then the simulations would not suggest dynamo action
from the $\meanWW\times\meanJJ$ effect.
In any case, it is important to consider simulations with
larger magnetic Reynolds.
So far there is only the example of solar-like shear flow turbulence of
Brandenburg (2005) where the $\meanWW\times\meanJJ$ effect is a leading
candidate for explaining the generation of large scale fields in the
absence of helicity.

\section{Conclusions}

The present investigations have demonstrated a number of new and
previously unknown properties both for forced and MRI-driven turbulence
in the shearing box approximation.
In many respects these two cases are quite different.
First of all, the fact that for MRI-driven turbulence most of
the dissipation happens near the upper and lower boundaries of the
disc (i.e.\ in the disc corona) is peculiar to this case and is
not in general true.
Indeed, for non-magnetic discs most of the dissipation is expected
near the midplane where the density is largest.
Without entering the discussion about the reality of non-magnetic
turbulence in accretion discs (e.g.\ Balbus et al.\ 1996), we note
that under some circumstances (e.g.\ in protostellar discs where the
electric conductivity is poor) the MRI is not likely to operate,
so less efficient mechanisms such as the inflow into the disc during
its formation and vertical shear (Urpin \& Brandenburg 1998) cannot
be excluded as possible agents facilitating turbulence.
Also the possibility of nonlinear instabilities (Richard \& Zahn 1999,
Chagelishvili et al.\ 2003, Afshordi et al.\ 2005) should be mentioned.
In any case, comparing magnetic and non-magnetic cases is important in
order to assess the potential validity of general turbulence concepts
that have mainly been studied under forced non-magnetic conditions.

There are a number of alternative ways of determining the turbulent
viscosity in discs.
Three methods have already been compared in the context of MRI-driven
turbulence by determining the stress either explicitly, from the heating
rate, or from the radial mass accretion rate (Brandenburg et al.\ 1996a).
A rather obvious alternative is to consider the decay of an initial
wave-like perturbation and to measure the decay of this signal.
Although this method gives sensible results in the context of
non-shearing and non-rotating flows (Yousef et al.\ 2003), in the
present case it cannot be used for this purpose,
because epicyclic oscillations are being initialized that hardly decay.
In fact, at least in the hydrodynamic case with not too strong forcing
there is clear evidence that epicyclic oscillations can actually be
excited stochastically--very much like the p-modes in the sun.
This may be important for understanding the quasi-periodic oscillations
discovered recently in some pulsars (e.g., Lee et al.\ 2004), and in
particular the connection with the possibility of exciting resonances
at the beat frequencies for non-equal vertical and radial epicyclic
frequencies.

Given that the turbulence is in general non-isotropic (owing to shear and
rotation, as well as stratification), the turbulent transport coefficients
are in general also non-isotropic.
For practical calculations of toroidally averaged accretion flows
(e.g., Kley et al.\ 1993, Igumenshchev et al.\ 1996) it is therefore
essential to know the full tensorial structure as well as
other possibly non-diffusive contributions.
Significant progress has been made in determining the full
tensorial structure of the turbulent resistivity and alpha effect.
Somewhat surprisingly, it turned out that the two diagonal components
of the resistivity tensor are nearly equal.
Furthermore, in the non-MRI case the off-diagonal components can even
given rise to dynamo action (the so-called $\meanWW\times\meanJJ$ or
shear--current effect).
As far as the $\alpha$ effect is concerned, an earlier result about the
different signs for MRI and forced turbulence is confirmed.

Similar investigations can probably also be carried out for determining
the tensorial nature of turbulent viscosity and possibly other
non-diffusive effects that are known in other circumstances
(e.g., R\"udiger \& Hollerbach 2004).
Most important is perhaps the investigation of stochastically
excited epicyclic oscillations in connection with the kilohertz
quasiperiodic oscillations found in some pulsars.
Obviously, more systematic investigations should be carried out to
determine the dependence on forcing amplitude of the turbulence and
to check whether similar oscillations are also possible for
MRI-driven turbulence.

\acknowledgements
I thank Eric G.\ Blackman, G\"unther R\"udiger, and Kandaswamy Subramanian
for suggestions and comments on the manuscript,
and Karl-Heinz R\"adler and Martin Schrinner for collaborating with
me on the implementation and testing of the technique described in
\Sec{Tensorial}.
The Danish Center for Scientific Computing is acknowledged for granting
time on the Horseshoe cluster.

\appendix
\section{The forcing function}
\label{ForcingFunction}

For completeness we specify here the forcing function used in the
present paper\footnote{This forcing function was also used by
Brandenburg (2001), but in his Eq.~(5) the factor 2 in the denominator
should have been replaced by $\sqrt{2}$ for a proper normalization.}.
It is defined as
\EQ
\ff(\xx,t)={\rm Re}\{N\ff_{\kk(t)}\exp[\ii\kk(t)\cdot\xx+\ii\phi(t)]\},
\EN
where $\xx$ is the position vector.
The wavevector $\kk(t)$ and the random phase
$-\pi<\phi(t)\le\pi$ change at every time step, so $\ff(\xx,t)$ is
$\delta$-correlated in time.
For the time-integrated forcing function to be independent
of the length of the time step $\delta t$, the normalization factor $N$
has to be proportional to $\delta t^{-1/2}$.
On dimensional grounds it is chosen to be
$N=f_0 c_{\rm s}(|\kk|c_{\rm s}/\delta t)^{1/2}$, where $f_0$ is a
nondimensional forcing amplitude.
The value of the coefficient $f_0$ is chosen such that the maximum Mach
number stays below about 0.5; in practice this means $f_0=0.01\ldots0.05$,
depending on the average forcing wavenumber.
At each timestep we select randomly one of many possible wavevectors
in a certain range around a given forcing wavenumber.
The average wavenumber is referred to as $k_{\rm f}$.
Two different wavenumber intervals are considered: $1...2$ for
$k_{\rm f}=1.5$ and $4.5...5.5$ for $k_{\rm f}=5$.
We force the system with transverse helical waves,
\EQ
\ff_{\kk}=
\left(\kk\times\eee\right)/\sqrt{\kk^2-(\kk\cdot\eee)^2},
\label{nohel_forcing}
\EN
where $\eee$ is an arbitrary unit vector
not aligned with $\kk$; note that $|\ff_{\kk}|^2=1$.

\section{Dispersion relation for tensorial $\tilde\eta$}
\label{DisperRel}

The dispersion relation for the dynamo problem with $\delta$ effect is
(Brandenburg \& Subramanian 2005)
\EQ
\lambda_\pm=-\eta_{\rm T} k^2
\pm\sqrt{-(\kk\cdot\ddelta)^2 k^2-(\kk\cdot\ddelta)Sk_z},
\label{lam_deltaOmega}
\EN
where $\lambda$ is the growth rate, $k_x$ and $k_z$ are the wavenumbers
in the $x$ and $z$ directions, $\kk^2=k_x^2+k_z^2$,
and $\eta_{\rm T}=\eta+\eta_{\rm t}$ the sum of microscopic
and turbulent diffusion.

We now consider the one-dimensional problem ($k_x=0$) and
treat all four components of $\tilde\eta_{ij}$ as independent,
so we have the following set of two partial differential equations,
\EQ
\dot{\meanB}_x=(\eta+\tilde\eta_{yy})\meanB_x''-\tilde\eta_{yx}\meanB_y'',
\EN
\EQ
\dot{\meanB}_y=(\eta+\tilde\eta_{xx})\meanB_y''-\tilde\eta_{xy}\meanB_x''+S\meanB_x,
\EN
where $S=-{3\over2}\Omega$ is the shear coefficient.
Using the abbreviation $n_{ij}=(\eta\delta_{ij}+\tilde\eta_{ij})k_z^2$,
this leads to the dispersion relation
\EQ
\lambda_\pm=-\half(n_{xx}\!+n_{yy})
\pm\sqrt{\quarter(n_{xx}\!-\!n_{yy})^2+n_{yx}(n_{xy}\!+\!S)},
\label{disper2}
\EN
which reduces to the one-dimensional form of \Eq{lam_deltaOmega} when
$\tilde\eta_{xx}=\tilde\eta_{yy}=\eta_{\rm t}$ and
$\tilde\eta_{xy}=-\tilde\eta_{yx}=\delta$.

\vfill\bigskip\noindent\tiny\begin{verbatim}
$Header: /home/brandenb/CVS/tex/disc/QPO/paper.tex,v 1.53 2005/10/01 08:36:11 brandenb Exp $
\end{verbatim}


\begin{thebibliography}{}

\bibitem{}
Abramowicz, M. A., Bulik, T., Bursa, M., Klu\'zniak, W.\yana{2003a}{404}{L21}

\bibitem{}
Abramowicz, M. A., Karas, V., Klu\'zniak, W., Lee, W. H.,
Rebusco, P.\ypasj{2003b}{55}{467}

\bibitem{}
Afshordi, N., Mukhopadhyay, B., Narayan, R.\yapj{2005}{629}{373}

\bibitem{}
Arlt R., R\"udiger G.\yana{2001}{374}{1035}

\bibitem{}
Balbus, S. A.\yapj{2004}{600}{865}

\bibitem{}
Balbus, S. A. Hawley, J. F.\yapj{1991}{376}{214}

\bibitem{}
Balbus, S. A. Hawley, J. F.\yjour{1998}{ReMP}{70}{1}

\bibitem{}
Balbus, S. A., Papaloizou, J. C. B.\yapj{1999}{521}{650}

\bibitem{}
Balbus, S. A., Hawley, J. F., Stone, J. M.\yapj{1996}{467}{76}

\bibitem{}
Blackman, E. G.: 2005, personal communication

\bibitem{}
Brandenburg, A.\yproc{1998}{61}
{Theory of Black Hole Accretion Discs}
{M. A. Abramowicz, G. Bj\"ornsson, J. E. Pringle}
{Cambridge University Press, Cambridge}

\bibitem{}
Brandenburg, A.\yapj{2001}{550}{824} (B01)

\bibitem{}
Brandenburg, A.\yproc{2003}{269}
{Advances in nonlinear dynamos}
{A. Ferriz-Mas, M. N\'u\~nez}
{Taylor \& Francis, London}

\bibitem{}
Brandenburg, A.\yapj{2005}{625}{539}

\bibitem{}
Brandenburg, A., Sokoloff, D.\ygafd{2002}{96}{319}

\bibitem{}
Brandenburg, A., Subramanian, K.\yan{2005a}{326}{400}

\bibitem{}
Brandenburg, A., Subramanian, K.\yjour{2005b}{PhR}{417}{1}

\bibitem{}
Brandenburg, A., Nordlund, \AA., Stein, R. F.,
Torkelsson, U.\yapj{1995}{446}{741}

\bibitem{}
Brandenburg, A., Nordlund, \AA., Stein, R. F.,
Torkelsson, U.\yapjl{1996a}{458}{L45} 

\bibitem{}
Brandenburg, A., Nordlund, \AA., Stein, R. F., Torkelsson, U.\yproc{1996b}{285}
{Physics of Accretion Disks}
{S. Kato, S. Inagaki, S. Mineshige, J. Fukue}
{Gordon and Breach Science Publishers}

\bibitem{}
Brandenburg, A., Dintrans, B., Haugen, N. E. L.\yproc{2004}{122}
{MHD Couette flows: experiments and models}
{R. Rosner, G. R\"udiger, A. Bonanno}
{AIP Conf. Proc. {\bf 733}}

\bibitem{}
Carballido, A., Stone, J. M., Pringle, J. E.\ymn{2005}{358}{1055}

\bibitem{}
Chagelishvili, G. D., Zahn, J.-P., Tevzadze, A. G.,
Lominadze, J. G.\yana{2003}{402}{401}

\bibitem{}
De Villiers, J.-P., Hawley, J. F.\yapj{2003}{592}{1060}

\bibitem{}
Frank, J., King, A. R., \& Raine, D. J.\ybook{1992}{Accretion power
in astrophysics}{Cambridge University Press, Cambridge}

\bibitem{}
Goldreich, P., Kumar, P.\yapj{1990}{363}{694}

\bibitem{}
Hawley, J. F., Gammie, C. F., Balbus, S. A.\yapj{1995}{440}{742}

\bibitem{}
Hawley, J. F., Gammie, C. F., Balbus, S. A.\yapj{1996}{464}{690}

\bibitem{}
Hawley, J. F.\yapj{2000}{528}{462}

\bibitem{}
Igumenshchev, I. V., Chen, X., Abramowicz, M. A.\ymn{1996}{278}{236}

\bibitem{}
Johansen, A., Klahr, H.\papj{2005}
(arXiv: astro-ph/0501641)

\bibitem{}
Kato, Y.\ypasj{2004}{56}{931}

\bibitem{}
Kley, W., Papaloizou, J. C. B., Lin, D. N. C.\yapj{1993}{409}{739}

\bibitem{}
Kowal, G., Otmianowska-Mazur, K., Hanasz, M.\yproc{2005}{171}
{The magnetized plasma in galaxy evolution}
{K.T. Chy\.zy, K. Otmianowska-Mazur, M. Soida, and R.-J. Dettmar}
{Jagiellonian University}

\bibitem{}
Klu\'zniak, W., Abramowicz, M. A., Kato, S., Lee, W. H.,
Stergioulas, N.\yapj{2004}{603}{L89}

\bibitem{}
Krause, F., R\"adler, K.-H.\ybook{1980}
{Mean-Field Magneto\-hydro\-dy\-na\-mics and Dynamo Theory}
{Pergamon Press, Oxford}

\bibitem{}
Lee, W. H., Abramowicz, M. A., Klu\'zniak, W.\yapj{2004}{603}{L93}

\bibitem{}
Matsumoto, R., Tajima, T.\yapj{1995}{445}{767}

\bibitem{}
Parker, E. N.\ybook{1979}{Cosmical Magnetic Fields}{Clarendon Press, Oxford}

\bibitem{}
R\"adler, K.-H.\yjour{1969}{Geod. Geophys. Ver\"off., Reihe II}{13}{131}

\bibitem{}
R\"adler, K.-H., Stepanov, R.\sjfm{2005}

\bibitem{}
Richard, D., Zahn, J.-P.\yana{1999}{347}{734}

\bibitem{}
Roberts, P. H., Soward, A. M.\yan{1975}{296}{49}

\bibitem{}
Rogachevskii, I., Kleeorin, N.\ypre{2003}{68}{036301}

\bibitem{}
Rogachevskii, I., Kleeorin, N.\ypre{2004}{70}{046310}

\bibitem{}
R\"udiger, G.\yact{1987}{37}{223}

\bibitem{}
R\"udiger, G., Pipin, V. V.\yana{2000}{362}{756}

\bibitem{}
R\"udiger, G., Hollerbach, R.\ybook{2004}{The magnetic universe}
{Wiley-VCH, Weinheim} 

\bibitem{}
R\"udiger, G., Kitchatinov, L. L.\spre{2005}

\bibitem{}
Schrinner, M., R\"adler, K.-H., Schmitt, D., Rheinhardt, M.,
Christensen, U.\yan{2005}{326}{245}

\bibitem{}
Shakura, N. I., Sunyaev, R. A.\yana{1973}{24}{337}

\bibitem{}
Stone, J. M., Hawley, J. F., Gammie, C. F.,
Balbus, S. A.\yapj{1996}{463}{656}

\bibitem{}
Torkelsson, U., Ogilvie, G. I., Brandenburg, A., Pringle, J. E.,
Nordlund, \AA., \& Stein, R. F.\ymn{2000}{318}{47}

\bibitem{}
Turner, N. J.\yapj{2004}{605}{L45}

\bibitem{}
Urpin, V., Brandenburg, A.\ymn{1998}{294}{399}

\bibitem{}
Yousef, T. A., Brandenburg, A., R\"udiger, G.\yana{2003}{411}{321}

\end{thebibliography}
\end{document}